\makeatletter

\font\tenmsx=msxm10
\font\sevenmsx=msxm7
\font\fivemsx=msxm5
\font\tenmsy=msym10
\font\sevenmsy=msym7
\font\fivemsy=msym5
\newfam\msxfam
\newfam\msyfam
\textfont\msxfam=\tenmsx  \scriptfont\msxfam=\sevenmsx
  \scriptscriptfont\msxfam=\fivemsx
\textfont\msyfam=\tenmsy  \scriptfont\msyfam=\sevenmsy
  \scriptscriptfont\msyfam=\fivemsy

\def\hexnumber@#1{\ifnum#1<10 \number#1\else
 \ifnum#1=10 A\else\ifnum#1=11 B\else\ifnum#1=12 C\else
 \ifnum#1=13 D\else\ifnum#1=14 E\else\ifnum#1=15 F\fi\fi\fi\fi\fi\fi\fi}

\def\msx@{\hexnumber@\msxfam}
\def\msy@{\hexnumber@\msyfam}
\mathchardef\boxdot="2\msx@00
\mathchardef\boxplus="2\msx@01
\mathchardef\boxtimes="2\msx@02
\mathchardef\square="0\msx@03
\mathchardef\blacksquare="0\msx@04
\mathchardef\centerdot="2\msx@05
\mathchardef\lozenge="0\msx@06
\mathchardef\blacklozenge="0\msx@07
\mathchardef\circlearrowright="3\msx@08
\mathchardef\circlearrowleft="3\msx@09
\mathchardef\rightleftharpoons="3\msx@0A
\mathchardef\leftrightharpoons="3\msx@0B
\mathchardef\boxminus="2\msx@0C
\mathchardef\Vdash="3\msx@0D
\mathchardef\Vvdash="3\msx@0E
\mathchardef\vDash="3\msx@0F
\mathchardef\twoheadrightarrow="3\msx@10
\mathchardef\twoheadleftarrow="3\msx@11
\mathchardef\leftleftarrows="3\msx@12
\mathchardef\rightrightarrows="3\msx@13
\mathchardef\upuparrows="3\msx@14
\mathchardef\downdownarrows="3\msx@15
\mathchardef\upharpoonright="3\msx@16

\mathchardef\downharpoonright="3\msx@17
\mathchardef\upharpoonleft="3\msx@18
\mathchardef\downharpoonleft="3\msx@19
\mathchardef\rightarrowtail="3\msx@1A
\mathchardef\leftarrowtail="3\msx@1B
\mathchardef\leftrightarrows="3\msx@1C
\mathchardef\rightleftarrows="3\msx@1D
\mathchardef\Lsh="3\msx@1E
\mathchardef\Rsh="3\msx@1F
\mathchardef\rightsquigarrow="3\msx@20
\mathchardef\leftrightsquigarrow="3\msx@21
\mathchardef\looparrowleft="3\msx@22
\mathchardef\looparrowright="3\msx@23
\mathchardef\circeq="3\msx@24
\mathchardef\succsim="3\msx@25
\mathchardef\gtrsim="3\msx@26
\mathchardef\gtrapprox="3\msx@27
\mathchardef\multimap="3\msx@28
\mathchardef\therefore="3\msx@29
\mathchardef\because="3\msx@2A
\mathchardef\doteqdot="3\msx@2B

\mathchardef\triangleq="3\msx@2C
\mathchardef\precsim="3\msx@2D
\mathchardef\lesssim="3\msx@2E
\mathchardef\lessapprox="3\msx@2F
\mathchardef\eqslantless="3\msx@30
\mathchardef\eqslantgtr="3\msx@31
\mathchardef\curlyeqprec="3\msx@32
\mathchardef\curlyeqsucc="3\msx@33
\mathchardef\preccurlyeq="3\msx@34
\mathchardef\leqq="3\msx@35
\mathchardef\leqslant="3\msx@36
\mathchardef\lessgtr="3\msx@37
\mathchardef\backprime="0\msx@38
\mathchardef\risingdotseq="3\msx@3A
\mathchardef\fallingdotseq="3\msx@3B
\mathchardef\succcurlyeq="3\msx@3C
\mathchardef\geqq="3\msx@3D
\mathchardef\geqslant="3\msx@3E
\mathchardef\gtrless="3\msx@3F
\mathchardef\sqsubset="3\msx@40
\mathchardef\sqsupset="3\msx@41
\mathchardef\trianglerighteq="3\msx@44
\mathchardef\trianglelefteq="3\msx@45
\mathchardef\bigstar="0\msx@46
\mathchardef\between="3\msx@47
\mathchardef\blacktriangledown="0\msx@48
\mathchardef\blacktriangleright="3\msx@49
\mathchardef\blacktriangleleft="3\msx@4A
\mathchardef\blacktriangle="0\msx@4E
\mathchardef\triangledown="0\msx@4F
\mathchardef\eqcirc="3\msx@50
\mathchardef\lesseqgtr="3\msx@51
\mathchardef\gtreqless="3\msx@52
\mathchardef\lesseqqgtr="3\msx@53
\mathchardef\gtreqqless="3\msx@54
\mathchardef\Rrightarrow="3\msx@56
\mathchardef\Lleftarrow="3\msx@57
\mathchardef\veebar="2\msx@59
\mathchardef\barwedge="2\msx@5A
\mathchardef\doublebarwedge="2\msx@5B
\mathchardef\angle="0\msx@5C
\mathchardef\measuredangle="0\msx@5D
\mathchardef\sphericalangle="0\msx@5E
\mathchardef\varpropto="3\msx@5F
\mathchardef\smallsmile="3\msx@60
\mathchardef\smallfrown="3\msx@61
\mathchardef\Subset="3\msx@62
\mathchardef\Supset="3\msx@63
\mathchardef\Cup="2\msx@64

\mathchardef\Cap="2\msx@65

\mathchardef\curlywedge="2\msx@66
\mathchardef\curlyvee="2\msx@67
\mathchardef\leftthreetimes="2\msx@68
\mathchardef\rightthreetimes="2\msx@69
\mathchardef\subseteqq="3\msx@6A
\mathchardef\supseteqq="3\msx@6B
\mathchardef\bumpeq="3\msx@6C
\mathchardef\Bumpeq="3\msx@6D
\mathchardef\lll="3\msx@6E

\mathchardef\ggg="3\msx@6F

\mathchardef\circledS="0\msx@73
\mathchardef\pitchfork="3\msx@74
\mathchardef\dotplus="2\msx@75
\mathchardef\backsim="3\msx@76
\mathchardef\backsimeq="3\msx@77
\mathchardef\complement="0\msx@7B
\mathchardef\intercal="2\msx@7C
\mathchardef\circledcirc="2\msx@7D
\mathchardef\circledast="2\msx@7E
\mathchardef\circleddash="2\msx@7F
\def\ulcorner{\delimiter"4\msx@70\msx@70 }
\def\urcorner{\delimiter"5\msx@71\msx@71 }
\def\llcorner{\delimiter"4\msx@78\msx@78 }
\def\lrcorner{\delimiter"5\msx@79\msx@79 }
\def\yen{\mathhexbox\msx@55 }
\def\checkmark{\mathhexbox\msx@58 }
\def\circledR{\mathhexbox\msx@72 }
\def\maltese{\mathhexbox\msx@7A }
\mathchardef\lvertneqq="3\msy@00
\mathchardef\gvertneqq="3\msy@01
\mathchardef\nleq="3\msy@02
\mathchardef\ngeq="3\msy@03
\mathchardef\nless="3\msy@04
\mathchardef\ngtr="3\msy@05
\mathchardef\nprec="3\msy@06
\mathchardef\nsucc="3\msy@07
\mathchardef\lneqq="3\msy@08
\mathchardef\gneqq="3\msy@09
\mathchardef\nleqslant="3\msy@0A
\mathchardef\ngeqslant="3\msy@0B
\mathchardef\lneq="3\msy@0C
\mathchardef\gneq="3\msy@0D
\mathchardef\npreceq="3\msy@0E
\mathchardef\nsucceq="3\msy@0F
\mathchardef\precnsim="3\msy@10
\mathchardef\succnsim="3\msy@11
\mathchardef\lnsim="3\msy@12
\mathchardef\gnsim="3\msy@13
\mathchardef\nleqq="3\msy@14
\mathchardef\ngeqq="3\msy@15
\mathchardef\precneqq="3\msy@16
\mathchardef\succneqq="3\msy@17
\mathchardef\precnapprox="3\msy@18
\mathchardef\succnapprox="3\msy@19
\mathchardef\lnapprox="3\msy@1A
\mathchardef\gnapprox="3\msy@1B
\mathchardef\nsim="3\msy@1C
\mathchardef\napprox="3\msy@1D
\mathchardef\nsubseteqq="3\msy@22
\mathchardef\nsupseteqq="3\msy@23
\mathchardef\subsetneqq="3\msy@24
\mathchardef\supsetneqq="3\msy@25
\mathchardef\subsetneq="3\msy@28
\mathchardef\supsetneq="3\msy@29
\mathchardef\nsubseteq="3\msy@2A
\mathchardef\nsupseteq="3\msy@2B
\mathchardef\nparallel="3\msy@2C
\mathchardef\nmid="3\msy@2D
\mathchardef\nshortmid="3\msy@2E
\mathchardef\nshortparallel="3\msy@2F
\mathchardef\nvdash="3\msy@30
\mathchardef\nVdash="3\msy@31
\mathchardef\nvDash="3\msy@32
\mathchardef\nVDash="3\msy@33
\mathchardef\ntrianglerighteq="3\msy@34
\mathchardef\ntrianglelefteq="3\msy@35
\mathchardef\ntriangleleft="3\msy@36
\mathchardef\ntriangleright="3\msy@37
\mathchardef\nleftarrow="3\msy@38
\mathchardef\nrightarrow="3\msy@39
\mathchardef\nLeftarrow="3\msy@3A
\mathchardef\nRightarrow="3\msy@3B
\mathchardef\nLeftrightarrow="3\msy@3C
\mathchardef\nleftrightarrow="3\msy@3D
\mathchardef\divideontimes="2\msy@3E
\mathchardef\varnothing="0\msy@3F
\mathchardef\nexists="0\msy@40
\mathchardef\mho="0\msy@66
\mathchardef\thorn="0\msy@67
\mathchardef\beth="0\msy@69
\mathchardef\gimel="0\msy@6A
\mathchardef\daleth="0\msy@6B
\mathchardef\lessdot="3\msy@6C
\mathchardef\gtrdot="3\msy@6D
\mathchardef\ltimes="2\msy@6E
\mathchardef\rtimes="2\msy@6F
\mathchardef\shortmid="3\msy@70
\mathchardef\shortparallel="3\msy@71
\mathchardef\smallsetminus="2\msy@72
\mathchardef\thicksim="3\msy@73
\mathchardef\thickapprox="3\msy@74
\mathchardef\approxeq="3\msy@75
\mathchardef\succapprox="3\msy@76
\mathchardef\precapprox="3\msy@77
\mathchardef\curvearrowleft="3\msy@78
\mathchardef\curvearrowright="3\msy@79
\mathchardef\digamma="0\msy@7A
\mathchardef\varkappa="0\msy@7B
\mathchardef\hslash="0\msy@7D
\mathchardef\hbar="0\msy@7E
\mathchardef\backepsilon="3\msy@7F
\def\Bbb{\ifmmode\let\next\Bbb@\else
 \def\next{\errmessage{Use \string\Bbb\space only in math mode}}\fi\next}
\def\Bbb@#1{{\Bbb@@{#1}}}
\def\Bbb@@#1{\fam\msyfam#1}

\def\inv{^{\raise.15ex\hbox{${
  \scriptscriptstyle -}$}\kern-.05em 1}}

\def\Dsl{\,\raise.15ex\hbox{$/$}\mkern-13.5mu D}
\def\dsl{\raise.15ex\hbox{$/$}\kern-.57em\hbox{$\partial$}}

\def\lspace{\ifx\answ\bigans{}\else\qquad\fi}
\def\del{\partial}
\def\CR{\hbox{{$\cal R$}}}

\def\lform{\hbox{$\sqcup$}\llap{\hbox{$\sqcap$}}}
\def\darr#1{\raise1.5ex\hbox{$\leftrightarrow$}
\mkern-16.5mu #1}
\def\INT{{\textstyle \int\kern-.642em\int}}

\def\R{{\Bbb R}}
\def\C{{\Bbb C}}
\def\Z{{\Bbb Z}}

\def\eps{{\epsilon}}

\def\cocross{{>\!\!\!\triangleleft}}

\def\tens{\mathop{\otimes}}

\def\la{{\triangleright}}\def\ra{{\triangleleft}}

\def\id{{\rm id}}
\def\Deltaop{{\Delta^{\rm op}}}

\def\nquad{{\!\!\!\!\!\!}}
\def\nqquad{\nquad\nquad}
\def\eqn#1#2{\begin{equation}#2\label{#1}\end{equation}}
\def\o{{}_{(1)}}\def\t{{}_{(2)}}\def\th{{}_{(3)}}

\def\bo{{}^{\bar{(1)}}}\def\bt{{}^{\bar{(2)}}}

\def\und#1{{\underline {#1}}}

\def\uo{{{}^{(1)}}}\def\ut{{{}^{(2)}}}

\def\new#1{\goodbreak\goodbreak\bigskip
\noindent{\bf #1}}

\def\text#1{\mbox{\rm #1}}
\def\note#1{}

\def\blacksquare{{\lform}}%AMS Tex Fakes
\def\frac#1#2{{{#1\over#2}}}

\def\proof{\goodbreak\noindent{\bf Proof\quad}}
\def\endproof{{\ $\lform$}\bigskip }

\def\align#1{\begin{eqnarray*}#1\end{eqnarray*}}
\def\und#1{{\underline{#1}}}

\def\Bo{{{}_{\und{(1)}}}}\def\Bt{{{}_{\und{(2)}}}}
\def\vect{{\bf t}}\def\vecs{{\bf s}}
\def\vecx{{\bf x}}\def\vecl{{\bf l}}
\def\vecp{{\bf p}}

\def\<{\langle}
\def\>{\rangle}

\makeatother

\documentstyle[11pt]{article}

\def\thebibliography#1{\section*{REFERENCES}\list
 {[\arabic{enumi}]}{\settowidth\labelwidth{[#1]}\leftmargin\labelwidth
 \advance\leftmargin\labelsep
 \usecounter{enumi}}
 \def\newblock{\hskip .11em plus .33em minus -.07em}
 \sloppy
 \sfcode`\.=1000\relax}

\newtheorem{lemma}{Lemma}
\newtheorem{propos}[lemma]{Proposition}
\newtheorem{example}[lemma]{Example}
\newtheorem{theorem}[lemma]{Theorem}

\newtheorem{corol}[lemma]{Corollary}

\begin{document}\baselineskip 25pt

{\ }\hskip 4.7in DAMTP/92-65
\vspace{.5in}

\begin{center} {\Large BRAIDED MOMENTUM IN THE Q-POINCARE GROUP}
\baselineskip 13pt{\ }\\
{\ }\\ Shahn Majid\footnote{SERC Fellow and Fellow of Pembroke College,
Cambridge}\\ {\ }\\
Department of Applied Mathematics\\
\& Theoretical Physics\\ University of Cambridge\\ Cambridge CB3 9EW, U.K.
\end{center}

\begin{center}
October 1992\end{center}
\vspace{10pt}
\begin{quote}\baselineskip 13pt

\centerline{\bf ABSTRACT}
The $q$-Poincar\'e group of \cite{SWW:inh} is shown to have the structure of a
semidirect product and coproduct $B\cocross \widetilde{SO_q(1,3)}$ where $B$ is
a braided-quantum group structure on the $q$-Minkowski space of 4-momentum with
braided-coproduct $\und\Delta \vecp=\vecp\tens 1+1\tens \vecp$. Here the
necessary $B$ is not a usual kind of quantum group, but one with braid
statistics. Similar braided-vectors and covectors $V(R')$, $V^*(R')$ exist for
a general R-matrix. The abstract structure of the $q$-Lorentz group is also
studied.
\end{quote}

\baselineskip 22pt

\section{Introduction}

Many authors have considered the possibility of $q$-deforming the Poincar\'e
group as part of a general programme of $q$-deforming conventional quantum
field theory. Such a programme, if successful, would introduce a new
regularization parameter in physics\cite{Ma:reg} while preserving all usual
symmetries as $q$-symmetries. Reasonable candidates for the (non-commutative)
algebra of `functions' on $q$-Minkowski space and the $q$-Lorentz group
$SO_q(1,3)$ are known\cite{CWSSW:lor} but when it comes to the Pioncar\'e group
a problem seems to be that there are too many
possibilities\cite{OSWZ:def}\cite{LRNT:def}\cite{Luk-K:new}\cite{SWW:inh} if
one merely looks for a Hopf algebra or quantum group structure. One technique
in narrowing down the possibilities has been to look for a Hopf $*$-algebra
structure and/or to look for a universal $R$-matrix or quasitriangular
structure. Both attempts have not been fully successful.

Instead, we would like to develop here a different structural consideration,
namely that the Poincar\'e group should be built up  from the $q$-momentum and
$q$-Lorentz as a semidirect product (in analogy with the classical situation).
This tells us how the complicated generators and relations are built up in a
conceptual way from simple ones. Quantum groups such as $SO_q(1,3)$ can
certainly act on other quantum groups (such as a $q$-momentum) just as in the
classical case. Hence the problem is a natural and well-posed one, but again
one that has so far not proven possible in a full sense (not just for the
algebra but for coalgebra also) for any of the candidates above. We explain
first that there are general grounds that this is not possible unless the
$q$-momentum is a braided-quantum group\cite{Ma:bra}\cite{Ma:exa} rather than
an ordinary quantum one. These objects are more like supergroups or
super-quantum groups, but with bose-fermi statistics replaced by braid
statistics. We then show that one of the proposed $q$-Poincar\'e groups, namely
that in \cite{SWW:inh} has just such a structure. A similar situation may
prevail for \cite{OSWZ:def}\cite{LRNT:def}. One may also ask about the abstract
structure of the $q$-Lorentz group itself. For completeness, the paper
concludes with some remarks about this.

In more detail, a braided-quantum group is a Hopf algebra with braid
statistics\cite{Ma:bra}\cite{Ma:exa}. This means an algebra $B$ equipped with a
coproduct homomorphism $B\to B\und\tens B$ where $B\und\tens B$ is not the
usual tensor product algebra, but instead the tensor-product with braid
statistics\cite{Ma:bra}\cite{Ma:exa}
\eqn{btens}{(a\tens b)(c\tens d)=a\Psi(b\tens c)d}
where $\Psi$ is the braided-transposition. This braiding $\Psi$ can arise in a
variety of ways, but the one that concerns us comes from the universal
$R$-matrix of some background quantum group.

The general grounds for needing such objects for the Poincare group are as
follows. Firstly, the Poincare group should be a Hopf algebra containing the
Lorentz group. Thus $U_q(so(1,3))\subset U_q(p)$ where $p$ stands for the
Poincare group Lie algebra and we wish to find its deformed enveloping algebra
$U_q(p)$. In  nice cases we should also be able to project $U_q(p)\to
U_q(so(1,3))$ by setting the momentum generators to zero (classically, on can
do this because of the semidirect structure of $p$). The projection should of
course cover the inclusion. These are some minimal conditions that we might
expect for the relationship between these two Hopf algebras. However, there is
a theorem\cite[Prop. A.2]{Ma:skl} that if these maps are Hopf algebra maps then
$U_q(p)=B\cocross U_q(so(1,3))$, as an algebra and coalgebra, for some
braided-Hopf algebra $B$. This is a mathematical theorem that holds for any
pair of Hopf algebras with an inclusion/projection between them and is related
to a theorem of Radford\cite{Rad:str}. Thus the quantum version of the familiar
semidirect product situation forces us {\em a priori} into the braided setting,
at least for the momentum. The same theorem applies when we look in the dual
form for the $q$-Poincare function algebra $P_q$. Here the usual inclusion
becomes a projection $P_q\to SO_q(1,3)$ and can be expected to also cover a
Hopf algebra inclusion $SO_q(1,3)\subset P_q$. From this alone we conclude that
$P_q=B\cocross SO_q(1,3)$ for some braided-Hopf algebra $B$ of function algebra
type. Moreover, the braiding $\Psi$ will be induced by the action of the
universal $R$-matrix for $U_q(so(1,3))$ on the momentum and can therefore be
expected to be non-trivial. We will obtain Hopf algebras of exactly this
general type, the only further complication being a slight extension of the
$q$-Lorentz group by a central element $g$ needed for technical reasons to do
with the normalization of $R$-matrices.

\section{Braided-Covectors}

{}From the above general arguments, we are led to look for a braided-momentum
group $B$ rather than a usual quantum group. In the dual function algebra
description it should surely coincide as an algebra with the $q$-Minkowski
space of \cite{CWSSW:lor}. We will demonstrate this as well as a general
$R$-matrix construction of which this is an example.

The basic idea is that the function-algebra of the momentum group should be
nothing other than a braided-commutative (like super-commutative) plane, with
trivial linear coproduct. If $R$ is the $SO_q(1,3)$ matrix, the most naive
possibility is the Zamalodchikov algebra $\vecx_1\vecx_2=\cdot\Psi(\vecx_1\tens
\vecx_2)=\vecx_2\vecx_1R_{12}$, where the braiding is given by $R$. On the
other hand, we have explained in \cite{Ma:exa} that this naive notion is not
good when $\Psi^2\ne \id$ as is the case here. Instead, one needs some variant
$R'$ when defining the notion of braided-commutativity.

\begin{theorem} Let $R$ be an invertible matrix in $M_n\tens M_n$ obeying the
QYBE and suppose that $R'$ is another invertible matrix such that

(i) $R_{12}R_{13}R'_{23}=R'_{23}R_{13}R_{12}$,
$R_{23}R_{13}R'_{12}=R'_{12}R_{13}R_{23}$

(ii) $(PR+1)(PR'-1)=0$ where $P$ is the permutation matrix.

\noindent Then the {\em braided-covectors} $V^*(R')$ defined by generators
$1,x_i$ and relations
 $\vecx_2\vecx_1R'_{12}=\vecx_1\vecx_2$ forms a braided-bialgebra with
\[ \und\Delta x_i=x_i\tens 1+1\tens x_i,\qquad \Psi(\vecx_1\tens
\vecx_2)=\vecx_2\tens \vecx_1R_{12}.\]
extended multiplicatively with braid statistics. The counit is $\eps(x_i)=0$.
If also

(iii) $R_{21}R'_{12}=R'_{21}R_{12}$

\noindent then $V^*(R')$ is a braided-Hopf algebra with braided-antipode $\und
Sx_i=-x_i$ extended anti-multiplicatively with braid statistics.
\end{theorem}
\proof The definition is very similar to the braided-matrices $B(R)$ introduced
in \cite{Ma:exa} with $R,R'$ in the role of matrices $\Psi,\Psi'$ there.
Firstly, $V^*(R')$ is by definition an associative algebra. We use here and
throughout a standard compact notation where numerical suffices as in
$R_{12},\vecx_1,\vecx_2$ etc refer to the position in a matrix tensor product
while within each tensor factor we use a standard matrix notation (so
$\vecx_1\vecx_2=\vecx_2\vecx_1R'_{12}$ means
$x_jx_l=x_kx_iR'{}^i{}_j{}^k{}_l$). We have to check that $\Psi, \und\Delta,
\und S$ are well-defined when extended to products. Firstly, $\Psi$ extends to
tensor products according to the rules of a braiding ($R$ generates a braided
category), namely
\[ \Psi(\vecx_1\tens (\vecx_2\tens \vecx_3))=(\id\tens \Psi)(\Psi(\vecx_1\tens
\vecx_2)\tens \vecx_3)=\vecx_2\tens\Psi(\vecx_1\tens
\vecx_3)R_{12}=\vecx_2\tens \vecx_3\tens \vecx_1 R_{13}R_{12}\]
etc. The extension to products is then in such a way that $\Psi$ is functorial
with respect to the product, in the sense \[ \Psi(\vecx_1\tens
\vecx_2\vecx_3)=(\cdot\tens\id)\Psi(\vecx_1\tens (\vecx_2\tens
\vecx_3))=\vecx_2\vecx_3\tens \vecx_1 R_{13}R_{12}.\]
To see that this extension is well defined, we compute also
\[ \Psi(\vecx_1\tens
\vecx_3\vecx_2R'_{23})=(\cdot\tens\id)\Psi(\vecx_1\tens(\vecx_3\tens
\vecx_2))R'_{23}=\vecx_3\vecx_2\tens \vecx_1 R_{12}R_{13}R'_{23}\]
which is consistent with the relation $\vecx_2\vecx_3=\vecx_3\vecx_2R'_{23}$
given the first of conditions (i). Hence $\Psi(\vecx_1\tens (\ ))$ is a
well-defined map on the algebra $V^*(R')$. One can then compute in the same way
from functoriality that
\eqn{psiright}{ \Psi(\vecx_1\tens \vecx_2\vecx_3\cdots
\vecx_a)=\vecx_2\vecx_3\cdots \vecx_a\tens \vecx_1 R_{1n}\cdots R_{12}.}
Using this, we compute in a similar way
\align{&&\nqquad\Psi(\vecx_1\vecx_2\tens \vecx_3\cdots
\vecx_a)=(\id\tens\cdot)\Psi((\vecx_1\tens \vecx_2)\tens \vecx_3\cdots
\vecx_a)=(\id\tens\cdot)\Psi(\vecx_1\tens \vecx_3\cdots \vecx_a)\tens \vecx_2
R_{2a}\cdots R_{23}\\
&&=\vecx_3\cdots \vecx_a\tens \vecx_1\vecx_2 R_{1a}\cdots R_{13}R_{2a}\cdots
R_{23}=\vecx_3\cdots \vecx_a\tens \vecx_2\vecx_1 R'_{12}R_{1a}\cdots
R_{13}R_{2n}\cdots R_{23}\\
&&\nqquad\Psi(\vecx_2\vecx_1R'_{12}\tens \vecx_3\cdots
\vecx_a)=(\id\tens\cdot)\Psi(\vecx_2\tens \vecx_3\cdots \vecx_a)\tens \vecx_1
R_{1a}\cdots R_{13}R'_{12}\\
&&=\vecx_3\cdots \vecx_a\tens \vecx_2\vecx_1 R_{2a}\cdots R_{23}R_{1a}\cdots
R_{13}R'_{12}.}
Here $R'_{12}R_{1a}\cdots R_{13}R_{2a}\cdots R_{23}=R'_{12}R_{1a}R_{2a}\cdots
R_{13}R_{23}$ since matrices living in disjoint tensor factors commute. We can
then repeatedly use the second of (i) to move $R'_{12}$ to the right to arrive
at  $R_{2a}R_{1a}\cdots R_{23}R_{13}R'_{12}=R_{23}R_{1a}\cdots R_{13}R'_{12}$.
Hence $\Psi:V^*(R')\tens V^*(R')\to V^*(R')\tens V^*(R')$ is well defined and
functorial with respect to the product. It takes the form of a transfer matrix
analogous to expressions in the theory of quantum inverse scattering.

Next we extend $\und\Delta$ to products in such a way that it is a homomorphism
to the braided tensor product (\ref{btens}),
\align{\und\Delta \vecx_1\vecx_2&&=(\vecx_1\tens 1+1\tens \vecx_1)(\vecx_2\tens
1+1\tens \vecx_2)=\vecx_1\vecx_2\tens 1+1\tens \vecx_1\vecx_2+\vecx_1\tens
\vecx_2+\Psi(\vecx_1\tens \vecx_2)\\
&&=\vecx_1\vecx_2\tens 1+1\tens \vecx_1\vecx_2+\vecx_1\tens
\vecx_2+\vecx_2\tens \vecx_1R_{12}}
\align{\und\Delta \vecx_2\vecx_1R'_{12}&&=(\vecx_2\tens 1+1\tens
\vecx_2)(\vecx_1\tens 1+1\tens \vecx_1)R'_{12}\\
&&=\vecx_2\vecx_1R'_{12}\tens 1+1\tens \vecx_2\vecx_1R'_{12}+\vecx_2\tens
\vecx_1 R'_{12}+\vecx_1\tens \vecx_2 R_{21}R'_{12}.}
Hence for $\und\Delta$ to be well-defined we need $\vecx_1\tens
\vecx_2(R_{21}R'_{12}-1)=\vecx_2\tens \vecx_1(R_{12}-R'_{12})$ or
$R_{21}R_{12}'-1=P(R-R')$ which is condition (ii). Here $P$ is the usual
permutation matrix $\vecx_1\tens \vecx_2P=\vecx_1\tens \vecx_2$. It is trivial
to see that the braiding $\Psi$ is then functorial with respect to the
coproduct $\und\Delta:V^*(R')\to V^*(R')\und\tens V^*(R')$.

Finally, for a Hopf algebra in a braided category, the antipode is
braided-anti-multiplicative in the sense $\und S(ab)=\cdot\Psi(\und Sa\tens\und
Sb)$. We define $\und S$ on products of the generators in this way. Then
\align{\und S(\vecx_1\vecx_2)&&=\cdot\Psi(\und S\vecx_1\tens \und
S\vecx_2)=\vecx_2\vecx_1R_{12}=\vecx_1\vecx_2R_{21}'R_{12}\\
\und S(\vecx_2\vecx_1R_{12}')&&=\cdot\Psi(\und S\vecx_2\tens \und
S\vecx_1)R'_{12}=\vecx_1\vecx_2 R_{21}R'_{12}.}
Thus $\und S$ here is well-defined by condition (iii). More generally one can
compute likewise
\eqn{ant}{\und S(\vecx_1\cdots \vecx_a)=(-1)^a\vecx_a\cdots \vecx_1
R_{12}\cdots R_{1a} R_{23}\cdots R_{2a}\cdots R_{a-1\ a}.}
\endproof

\begin{example} If $R$ is a Hecke symmetry in the sense that it obeys
\[ (PR+1)(PR-q^2)=0\]
(for example the standard $R$-matrix for all $SL_q(n)$ after a suitable
scaling), then $R'=q^{-2}R$ obeys the above. Hence in this case the usual
Zamalodchikov algebra $V^*(R)$ is a braided-Hopf algebra with the linear
braided-coproduct.
\end{example}

For example, the standard bosonic quantum plane with relations $xy=q^{-1}yx$ is
a braided-Hopf algebra with
\[ \und\Delta x=x\tens 1+1\tens x,\quad \und\Delta y=y\tens 1+1\tens y,\quad
\und Sx=-x,\quad \und Sy=-y\]
\[ \Psi(x\tens x)=q^2 x\tens x,\quad \Psi(x\tens y)=q y\tens x,\quad
\Psi(y\tens y)=q^2 y\tens y \]
\[ \Psi(y\tens x)=q x\tens y+(q^2-1)y\tens x.\]
We call this the {\em braided-plane}. The sub-braided-Hopf algebras generated
by $x$ or $y$ alone are copies of the {\em braided-line} as introduced in
\cite{Ma:csta}.

\begin{example} If $R$ is ($q$ times) the R-matrix for $SO_q(1,3)$ in
\cite{CWSSW:lor}, we take
\[ R'=P+\mu P(PR-q^{-2})(PR-q^2).\]
In this case we call $V^*(R')$ the {\em (right handed) braided-Minkowski space}
$\und{\R^{1,3}}$. As an algebra, it coincides with a right-handed version of
the $q$-Minkowski space in \cite{CWSSW:lor}, equipped now with the linear
braided-coproduct.
\end{example}
\proof The Hecke-type relation for this $R$-matrix (in our normalization)
is\cite{CWSSW:lor} $(PR+1)(PR-q^{-2})(PR-q^2)=0$. Hence condition (ii) in
Theorem~1 is obeyed. Moreover, since $PR'$ is a polynomial in $PR$ it follows
at once that conditions (i),(iii) are obeyed. To see this, it is convenient to
write them in terms of $PR,PR'$: after renumbering they are
$(PR)_{23}(PR)_{12}(PR')_{23}=(PR')_{12}(PR)_{23}(PR)_{12}$ and
$(PR)(PR')=(PR')(PR)$. Since $PR'=PR$ and $PR'=\id$ obey these, it follows that
so does any polynomial in $PR,\id$.
Hence we have a braided-covector Hopf algebra by the theorem. The value of
$\mu\ne 0$ does not change the resulting algebra but can be chosen so that $R'$
is invertible. Note that $PR'-1$ here is proportional to the antisymmetrizer
$P_A$ in the approach of \cite{CWSSW:lor} so that $V^*(R)$ coincides with (a
right-handed version of) their $q$-Minkowski space algebra as defined by
$\vecx_1\vecx_2P_A=0$.
\endproof

Clearly the last proof works for any $R$-matrix obeying a polynomial identity
$f(PR)=0$. Writing the roots as $\lambda_1$ we have
$(PR-\lambda_1)(PR-\lambda_2)\cdots (PR-\lambda_m)=0$. Hence for each non-zero
root $\lambda_i$ we scale $R$ so that $(PR+1)\prod_{j\ne
i}(PR+{\lambda_j\over\lambda_i})=0$ and define $PR'=1+\mu\prod_{j\ne
i}(PR+{\lambda_j\over\lambda_i})$. Hence there is a braided-covector Hopf
algebra for each non-zero root of the functional equation for $PR$ (each
non-zero eigenvalue of $PR$).

Also clearly, there are plenty of variants of our construction based on
slightly different conventions. For example, for the same data $R,R'$ as in
Theorem~1, we can define $V(R')$ by generators $x^i$ (upper index) and
relations $\vecx_1\vecx_2=R'\vecx_2\vecx_1$. This forms a {\em braided-vector}
Hopf algebra with linear braided-coproduct, braided-antipode and braiding,
\eqn{vec}{\und\Delta\vecx=\vecx\tens 1+1\tens\vecx,\quad\und
S\vecx=-\vecx,\quad \Psi(\vecx_1\tens\vecx_2)=R_{12}\vecx_2\tens\vecx_1.}
The proof that this is a braided-Hopf algebra is strictly analogous to the
proof of Theorem~1.

What about the physical meaning of these braided-coproducts? Our interpretation
from the point of view of non-commutative or braided geometry is that it
corresponds to a group-structure on the non-commutative or braided space
expressing an analog of the addition of covectors (in analogy with the meaning
in the commutative case). Such a group structure means of course that we can
make translations. Thus $\und\Delta: V^*(R')\to V^*(R')\und\tens V^*(R')$ can
be viewed as a braided-coaction on one copy of $B$ by  another. If we denote
the generators of the second copy by $p_i$ to distinguish them from $x_i$, the
braided-coaction $\und\beta$ is
\eqn{coaction}{\und\beta(x_i)=x_i\tens 1+1\tens p_i,\quad{\rm i.e.},\quad
x_i'=x_i+p_i}
where the second expression is a more compact notation for the coaction. This
innocent expression nevertheless carries non-trivial information, for we can
differentiate it to obtain the `braided-vector-fields' corresponding to
translation.
For example, for the braided line (the 1-dimensional case) the coaction
$x'=x+p$ when we remember the braid statistics $\Psi(p\tens x)=q^2 x\tens p$
between the two copies of the braided line, gives
\eqn{dq}{(x'{}^m-x^m)p^{-1}|_{p=0}=((x+p)^m-x^m)p^{-1}|_{p=0}=(\sum
\left[{m\atop r}\right]_q x^r p^{m-r}-x^m)p^{-1}|_{p=0}=[m]_q x^{m-1}}
which is the usual $q$-derivative $\del_q$ of $x^m$. We used here the notation
$[m]_q={q^{2m}-1\over q^2-1}$ and the usual $q$-binomial theorem to expand
$(x\tens 1+1\tens p)^m$ (the two terms $q^{-2}$-commute in the braided tensor
product algebra). Thus the usual $q$-derivative
$\del_q(f)(x)={f(q^2x)-f(x)\over (q^2-1)x}$ is the generator of
braided-translations on the braided-line. From this point of view its
well-known Leibnitz rule $\del_q (fg)=(\del_q f)g+L_q(f)(\del_qg)$ where
$L_q(f)(x)=f(q^2x)$ should be viewed as analogous to the super-Leibnitz rule,
with the operator $L_q$ playing the role of the usual $\pm1$ factor.

This point of view will be developed further elsewhere. For our present
purposes we note only that our braided-coproduct $\und\Delta$ thus replaces in
our approach the study of differential calculus that characterizes the approach
to the $q$-Poincar\'e group in \cite{OSWZ:def}. We work with finite
translations rather than infinitesimal ones.

\section{$q$-Poincare Group as a Semidirect Product}

We have developed the braided-covectors as living in the braided category
generated by an $R$-matrix (which plays the role of $\pm1$ in the super case).
On the other hand, this category is basically the category of comodules of the
bialgebra algebra $A(R)$\cite{FRT:lie} with matrix generator $\vect$ and
relations $R\vect_1\vect_2=\vect_2\vect_1R$ and equipped with a certain dual
quasitriangular structure $\CR:A(R)\tens A(R)\to \C$. This obeys some obvious
axioms dual to those for a universal $R$-matrix in \cite{Dri}, and is given
here by $\CR(\vect_1\tens\vect_2)=R$ extended to products as a skew
bialgebra-bicharacter along the lines developed in an equivalent form in
\cite[Sec. 3]{Ma:qua}. Thus
\eqn{dquaA(R)l}{\CR(\vect_1\cdots\vect_a\tens\vect_{a+1})
=\CR(\vect_1\tens\vect_{a+1})\cdots\CR(\vect_a\tens\vect_{a+1})
=R_{1a+1}\cdots R_{aa+1}}
\eqn{dquaA(R)r}{\CR(\vect_1\tens\vect_2\cdots\vect_{a+1})
=\CR(\vect_{a+1}\tens\vect_1)\cdots\CR(\vect_2\tens\vect_1)
=R_{a+1\ 1}\cdots R_{21}}
while the expression for
$\CR(\vect_1\cdots\vect_a\tens\vect_{a+1}\cdots\vect_{a+b})$ is an array of
R-matrices cf\cite[Sec. 5.2]{Ma:qua}. See \cite{Ma:eul} where we also explain
how it leads to the braiding $\Psi$ in the category. If $V,W$ are any two right
comodules of a dual-quasitriangular Hopf algebra, written explicitly as $v\to
v'=\sum v\bo\tens v\bt$ and $w\to w'=\sum w\bo\tens w\bt$ say, then the
braiding is
\eqn{braiding}{\Psi_{V\tens W}(v\tens w)=\sum w\bo \tens v\bo \CR(v\bt\tens
w\bt).}
This formula is all that we need here from the general category theory (we will
give direct proofs of other relevant facts).

\begin{propos} $V^*(R')$ lives in the braided category of right
$A(R)$-comodules: all its structure maps are fully covariant under the right
coaction $V^*(R')\to V^*(R')\tens A(R)$ given by $\vecx\to \vecx'=\vecx\vect$
and the braiding $\Psi$ on $V^*(R')$ is the one inherited from this.
\end{propos}
\proof This is the meaning of condition (i) in Theorem~1, for this is implied
by and essentially implies the identity
\eqn{R'tt}{ R'_{12}\vect_1\vect_2=\vect_2\vect_1R'_{12}.}
Firstly, note that applying the fundamental and conjugate fundamental
representations $\rho^+_2(\vect_1)=R_{12}$ and $\rho^-_2(\vect_1)=R_{21}^{-1}$
to (\ref{R'tt}) gives exactly the conditions (i) in Theorem~1, while in the
reverse direction, repeatedly using the conditions (i) establishes (\ref{R'tt})
in all tensor powers of these representations and hence essentially corresponds
to (\ref{R'tt}) abstractly. Given this, we have
$\vecx_1\vect_1\vecx_2\vect_2=\vecx_1\vecx_1\vect_2\vect_2=\vecx_2\vecx_1
R'_{12}\vect_1\vect_2=\vecx_2\vecx_1\vect_2\vect_1R'_{12}=\vecx_2\vect_2\vecx_1
\vect_1R'_{12}$.  This means that the algebra is covariant. Covariance of the
coalgebra $\und\Delta$ is immediate on the generators and hence holds in
general because the braiding (used in extending $\und\Delta$ to products) is
covariant. Covariance in both cases means that the relevant structure maps
are intertwiners for the quantum group action, as we have explained in detail
in \cite{Ma:lin}.
Another approach is to verify (\ref{R'tt}) directly without worrying about the
reconstruction. For, example, it holds in Examples~2,~3 and any similar
examples just because $PR'$ is a function of $PR$ and $\id$, both of which
commute with $\vect_1\vect_2$. Finally, we check that the braiding on $V^*(R')$
is indeed the one that we have used in Theorem~1. Thus,  from (\ref{braiding})
we see that the matrix coaction induces the braiding
$\Psi(\vecx_1\tens\vecx_2)=\vecx_2\tens\vecx_1
\CR(\vect_1\tens\vect_2)=\vecx_2\tens\vecx_1 R_{12}$. \endproof

Our next task is to extend this point of view to the case where the bialgebra
$A(R)$ is replaced by a quantum group function algebra $A$ (such as $SL_q(n),
SO_q(n)$ etc). These are quotients of $A(R)$ by further determinant-type (and
other) relations. See \cite{FRT:lie} for the standard cases but note that the
procedure works more generally as we have explained in \cite[Sec. 3]{Ma:qua}.
In the standard cases we know the result is again dual quasitriangular since
the corresponding $U_q(g)$ is quasitriangular\cite{Dri}, but we showed in
\cite[Sec. 3.2.3]{Ma:qua} that the same situation prevails in the general case
where the quotienting relations and pairing are defined from a generic
$R$-matrix (we showed that $U(R)$ was formally quasitriangular as a map),
provided $R$ is normalized correctly. In the present setting it means that
$\CR$ in (\ref{dquaA(R)l})-(\ref{dquaA(R)r}) etc are not consistent with the
determinant-type and other relations of $A$ unless $R$ is correctly normalized
(the quantum-group normalization).

In the present paper the normalization of $R$ is fixed by the requirement (ii)
of Theorem~1 and it is {\em not} in general the quantum group normalization.
Hence on $A$ the dual quasitriangular structure takes the form
$\CR(\vect_1\tens\vect_2)=\lambda R_{12}$ where $\lambda$ is a constant that
takes us to the quantum group normalization (for the standard $R$-matrices it
is a power of $q$). Hence
$\CR(\vect_1\cdots\vect_a\tens\vect_{a+1}\cdots\vect_{a+b})$ acquires an extra
factor $\lambda^{ab}$.

This $\lambda$ then spoils the last part of the proof of Proposition~4. The
situation is that $V^*(R')$ does not in general live in the braided category of
representations of the dual quasitriangular Hopf algebra $A$. Instead, we must
extend $A$ slightly by adjoining a single invertible group-like element $g$
commuting with $\vect$. The coproduct is $\Delta g=g\tens g$. This extended
Hopf algebra $\tilde A$ is the tensor product of $A$ with the group algebra
$\C\Z$. We also define a dual quasitriangular structure on $\C\Z$ by
$\CR(g^a\tens g^b)=\lambda^{-ab}$ along the lines in \cite{Ma:csta} but in a
dual form. This extends to a dual quasitriangular structure $\CR:\tilde
A\tens\tilde A\to \C$ as the tensor product one.

\begin{propos} Let $A(R)$ be equipped with its initial dual quasitriangular
structure as in (\ref{dquaA(R)l})-(\ref{dquaA(R)r}), and $\tilde A$ with the
dual quasitriangular structure as described. There is a map of dual
quasitriangular bialgebras $A(R)\to \tilde A$ defined by $\vect\to \vect g$.
Moreover, $V^*(R)$ lives in the braided category of $\tilde A$-comodules with
coaction $\vecx\to \vecx'=\vecx\vect g$.
\end{propos}
\proof The first part follows from the skew bimultiplicativity property as
already exploited in (\ref{dquaA(R)l})-(\ref{dquaA(R)r}). For example,
$\CR(\vect_1 g\tens \vect_2
g)=\CR(g\tens\vect_2g)\CR(\vect_1\tens\vect_2g)=\CR(g\tens g)\CR(g\tens
\vect_1)\CR(\vect_1\tens g)\CR(\vect_1\tens\vect_2)=\lambda^{-1}\lambda
R_{12}=R_{12}$, which agrees with the value on the matrix generators of $A(R)$
as in (\ref{dquaA(R)l})-(\ref{dquaA(R)r}). For the second part, we note that
$V^*(R')$ is $\Z$-graded by the degree of $x_i$. Thus we have a coaction of
$\C\Z$ by $\vecx\to \vecx'=\vecx g$ (extending to products as an algebra
homomorphism so $\vecx_1\cdots\vecx_a\to\vecx_1\cdots\vecx_a g^a$). This
coaction measures the degree (or in physical terms, the scaling dimension) of
any homogeneous function of the $x_i$ and $V^*(R')$ is covariant under it.
Also, $A$ coacts covariantly by $\vecx\to\vecx\vect$ since $A$ is a quotient of
$A(R)$ and covariance under the latter was proven in the first part of
Proposition~4. Moreover the two coactions are compatible and together they give
the coaction of $\tilde A$ as stated. The final part of the proof of
Proposition~4 now goes through much as before with
$\Psi(\vecx_1\tens\vecx_2)=\vecx_2\tens\vecx_1\CR(\vect_1g\tens\vect_2g)
=\vecx_2\tens\vecx_1 R_{12}$ as required. \endproof

Throughout this paper, if we want to work with only the bialgebra $A(R)$ (or
its $GL(R)$ variant obtained by inverting some elements) then we do  not need
to introduce $g$ and can use Proposition~4. On the other hand, if we want to
work with quantum groups such as $SL_q(n)$, $SO_q(n)$ etc, we have to work with
their extensions by $g$ and use Proposition~5. We give the formulae in the
latter case, but for the simpler formulae for $A(R)$ the reader can simply set
$g=1,\lambda=1$ throughout.

For example, the braided-Minkowski space $\und{\R^{1,3}}$ lives in the
braided-category of $\widetilde{SO_q(1,3)}$-comodules, a fact which enters into
many constructions by forcing us to remember the braid statistics induced by
this (in the same way as we remember the anti-commuting nature of Grassmann
variables). It also means that $\widetilde{SO_q(1,3)}$ acts naturally on
$\und{\R^{1,3}}$ in a way that respects its structure as both an algebra and
coalgebra. The physical meaning for the coalgebra is that when this is used to
make translations as in (\ref{coaction}), the translation is
$\widetilde{SO_q(1,3)}$-covariant in the sense that one may first translate and
then rotate $\vecx'\to\vecx'\vect g$ or first rotate $\vecx$ and the
displacement $\vecp$ and then make the displacement.

Mathematically, this covariance means that one may make a semidirect product by
this coaction of $\tilde A$ and obtain necessarily a Hopf algebra (in analogy
with the classical situation for the Poincar\'e group). This semidirect product
construction applies to any Hopf algebra in a braided category of this
type\cite{Ma:bos}\cite{Ma:mec}, a process that we have introduced and called
{\em bosonization} because it turns a braided Hopf algebra into an ordinary
one.

\begin{theorem} Let $V^*(R')$ be a braided-covector space as in Theorem~1 and
Proposition~5. Its bosonization is the ordinary Hopf algebra
$V^*(R')\cocross \tilde A$. Explicitly, it has subalgebras $V^*(R')$, $\tilde
A$ with generators $\vecp$ and $\vect$ and cross-relations and coproduct
\[ \vecp g=\lambda^{-1}g\vecp,\
\vecp_1\vect_2=\lambda\vect_2\vecp_1R_{12},\qquad \Delta \vecp=\vecp\tens \vect
g+1\tens\vecp,\qquad \Delta \vect=\vect\tens\vect, \ \Delta g=g\tens g.\]
\end{theorem}
\proof  We denote the generators of $V^*(R')$ now by $p_i$ as they will play
the role of momentum. As explained in detail in  \cite[Lemma~4.4]{Ma:mec}, the
right coaction of $\tilde A$ is turned by $\CR$ into a right
action, which comes out as
$\vecp_1\ra\vect_2=\vecp_1\CR(\vect_1g\tens\vect_2)=\lambda\vecp_1 R_{12}$ and
$\vecp\ra g=\vecp\CR(\vect g\tens g)=\vecp\lambda^{-1}$. In fact, we have
already obtained a similar action to that of $\vect$ (in another context) in
\cite[Sec. 6.3]{Ma:qua} and one may check directly in the same way as there
that $V^*(R')$ is covariant, while covariance under the action of $g$ is
immediate. Given the covariant action, one may make a right-handed semidirect
product in a standard way. See \cite[Sec. 4]{Ma:mec} for the right-handed
conventions we use here. It is built on $V^*(R')\tens \tilde A$ with cross
relations $(\vecp\tens 1)(1\tens\vect)=\sum
(1\tens\vect\o)(\vecp\ra\vect\t\tens 1)$ where $\Delta\vect=\sum
\vect\o\tens\vect\t$ is the coproduct of $\tilde A$. The present matrix form
gives the result stated. Similarly for $(\vecp\tens 1)(1\tens g)=(1\tens
g)(\vecp\tens 1)\lambda^{-1}$. Even more straightforwardly (but in a perhaps
unfamiliar dual language) the coaction of $\tilde A$ allows us to make a cross
coproduct of the coalgebra of $V^*(R')$ by the right coaction of $\tilde A$ as
a right-handed version of the standard formulae as recalled in \cite[Sec.
6]{Ma:qua}. This gives $\Delta \vecp\tens 1=\sum (\vecp\Bo\bo\tens 1)\tens
(1\tens \vecp\Bo\bt)(\vecp\Bt\tens 1)$ which computes as stated. Here
$\und\Delta \vecp=\sum \vecp\Bo\tens\vecp\Bt$ is the braided-coproduct and
$\beta(\vecp)=\sum \vecp\bo\tens\vecp\bt=\vecp\tens\vect g$ is the coaction.
There is also necessarily an antipode built from the antipode of $\tilde A$ and
the braided-antipode. It is also uniquely determined from the bialgebra shown.
\endproof

This applies of course to $\und{\R^{1,3}}$. Its bosonization
$\und{\R^{1,3}}\cocross \widetilde{SO_q(1,3)}$ can be called the (right-handed)
$q$-Poincare group $P_q$.

There is also a left-handed version of the above constructions for which the
formulae can be obtained by a symmetry principle. The left-handed Minkowski
space is defined by $P_A\vecx_1\vecx_2=0$ or $R'\vecx_1\vecx_2=\vecx_2\vecx_1$
as in \cite{CWSSW:lor} and its bosonization gives the structure of the
inhomogeneous quantum group in \cite{SWW:inh}.
There are several other variants also possible. For example, the
Poincar\'e-type Hopf algebra based on the (right-handed) braided-vectors
$V(R')$ in (\ref{vec}) is

\begin{propos} The bosonization of $V(R')$ is a semidirect product Hopf algebra
$V(R')\cocross \tilde A$ with cross relations and coproduct
\[ \vecx g=\lambda g\vecx,\ \vecx_1\vect_2=\lambda^{-1}\vect_2
R^{-1}_{12}\vecx_1,\quad \Delta x^i=x^j\tens g^{-1}St^i{}_j+1\tens x^i,\quad
\Delta\vect=\vect\tens\vect,\ \Delta g=g\tens g.\]
\end{propos}
\proof Here $V(R')$ lives in the braided category of $\tilde A$-comodules by
the right coaction $\vecx\to g^{-1}\vect^{-1}\vecx$ as an element of
$V(R')\tens \tilde A$. Its bosonization follows the same steps as in the proof
of Theorem~6. Thus $\CR$ turns the right coaction of $\tilde A$ into a right
action
$\vecx_1\ra\vect_2=\CR(g^{-1}\vect_1^{-1}\tens\vect_2)\vecx_1=\lambda^{-1}
R^{-1}_{12}\vecx_1$ and $\vecx\ra g=\CR(g^{-1}\vect^{-1}\tens g)=\lambda\vecx$.
The cross relations then come out from $(\vecx_1\tens 1)(1\tens\vect_2)
=(1\tens \vect_2)(\vecx_1\ra\vect_2\tens 1)$ and similarly for $(\vecx\tens 1)
(1\tens g)$. The semidirect coproduct by the right coaction $\beta(x^i)
=\sum x^i\bo\tens x^i\bt=x^j\tens g^{-1}S t^i{}_j$ is then computed in the
standard way. Note that $\Deltaop$ (the opposite coproduct) gives an equally
good Hopf algebra which in the present example has the matrix form
\eqn{deltaop}{\Deltaop\vecx=\vecx\tens 1+ g^{-1}\vect^{-1}\tens\vecx.}
\endproof

Theorem~6 and its variants such as in Proposition~7 have many applications,
allowing us to develop the $q$-Poincar\'e group $P_q$ along standard lines for
a semidirect product. The most important is that it coacts covariantly on
$q$-Minkowski space:

\begin{corol} The semidirect product $V^*(R')\cocross \tilde A$ coacts
covariantly on $V^*(R')$. Denoting the generators of the latter by $x_i$, the
coaction is $\vecx'=\vecx\vect g + \vecp$.
\end{corol}
\proof This is ensured by a general feature of bosonization that the
(co)representations of the original braided-Hopf algebra correspond to the
usual (co)-representations of the bosonization. The theorem (for modules) is
\cite[Thm. 4.2]{Ma:bos}, which we use now in a dual form for comodules. The
result, which can also be verified directly in our present setting, is
$\beta(\vecx)=(\beta_{\tilde A}\tens\id)\und\beta(\vecx)=(\beta_{\tilde
A}\tens\id)(\vecx\tens 1+1\tens\vecp)=\vecx\tens1\tens\vect g+1\tens\vecp\tens
1$ where $\beta_{\tilde A}$ is the coaction of $\tilde A$ on $V^*(R')$. The
result is written compactly as stated. The general theory of bosonization
ensures not only that the result is a right coaction  but that it is an algebra
homomorphism ($V^*(R')\cocross \tilde A$ coacts covariantly). In our case, we
can see it directly as
\align{\vecx'_1\vecx'_2\nquad&&=(\vecx_1\vect_1 g+\vecp_1)(\vecx_2\vect_2
g+\vecp_2)=\vecx_1\vecx_2\vect_1\vect_2
g^2+\vecp_1\vecp_2+\vecp_1\vecx_2\vect_2 g+\vecx_1\vect_1 g\vecp_2\\
\vecx'_2\vecx'_1R'_{12}\nquad&&=(\vecx_2\vect_2 g+\vecp_2)(\vecx_1\vect_1
g+\vecp_1)R'_{12}=\vecx_2\vecx_1R'_{12}\vect_1\vect_2
g^2+\vecp_2\vecp_1R'_{12}+\vecx_2\vect_2 g\vecp_1 R'_{12}+\vecp_2\vecx_1\vect_1
gR'_{12}}
The first two terms are clearly equal (using Proposition~5 for the first term)
while $\vecp_2\vecx_1\vect_1 gR'_{12}=\vecx_1\vecp_2\vect_1
gR'_{12}=\vecx_1\vect_1 g\vecp_2R_{21}R'_{12}=\vecx_1\vect_1
g\vecp_2+\vecx_1\vect_1 g\vecp_2 P(R-R')=
\vecx_1\vect_1 g\vecp_2+\vecx_2\vect_2 g\vecp_1R_{12}-\vecx_2\vect_2
g\vecp_1R'_{12}$ using in turn the cross-relations in Theorem~6, the condition
(ii) in Theorem~1 and the action of the usual permutation $P$. Using again the
cross-relations in Theorem~6 we see that the expressions for $\vecx'_1\vecx'_2$
and $\vecx'_2\vecx'_1R'_{12}$ coincide. It is trivial to see that the linear
braided-coproduct $\und \Delta$ on $V^*(R)$ is likewise covariant. \endproof

Similarly in the other conventions $V(R')\cocross \tilde A$ etc. As a second
application, let us note that the dual of a semidirect product as an algebra
and coalgebra is again a semidirect product as an algebra and coalgebra (the
product and coproduct become interchanged). Details are in \cite{Ma:mec}. Hence
an immediate corollary of our approach is

\begin{corol} The dual of the $q$-Poincar\'e group in Theorem~6 or in
\cite{SWW:inh} is also a semidirect product as an algebra and coalgebra
$B\cocross \tilde H$ where $B$ is dual to $V^*(R')$ and $\tilde H$ dual to
$\tilde A$.
\end{corol}
\proof This follows at once by dualizing the algebras and coalgebras in
Theorem~6 in a standard way. Another more conceptual route to the same result
is to dualise $V^*(R')$ in the braided category and bosonize the resulting
braided-Hopf algebra $B$ according to the original bosonization formulae in
\cite[Thm. 4.1]{Ma:bos}. Here the categorical dual has the opposite algebra and
opposite coalgebra to the usual dual, but has the advantage that it remains a
Hopf algebra in the same braided category. The resulting $B$ is then viewed as
a Hopf algebra in the category of left $\tilde H$-modules (rather than right
$\tilde A$-comodules) and the bosonized algebra is the semidirect product by
this action of $\tilde H$ and the bosonized coalgebra the semidirect product by
the left coaction $\beta(b)=\CR_{21}\la b$. \endproof

For example, we obtain
\eqn{envpoin}{U_q(p)=B\cocross \widetilde{U_q(so(1,3))}}
as a well-defined Hopf algebra, where $B$ is the braided-momentum group dual to
$\und{\R^{1,3}}$ in Example~3 and $\widetilde{U_q(so(1,3))}$ is the quantum
group dual to $\widetilde{SO_q(1,3)}$, acting by $q$-rotations and scaling. As
noted in the proof, there are two ways to compute this dualization. The first
is to dualise the algebra and coalgebra of $\und{\R^{1,3}}$ and of
$\und{\R^{1,3}}\cocross \widetilde{SO_q(1,3)}$ a standard way, while another
more natural way is to dualise $\und{\R^{1,3}}$ in the category and then
bosonize it.  Details will be given elsewhere. It should also be mentioned that
this $U_q(p)$ does not appear to be a Hopf $*$-algebra in the sense of
\cite{Wor:com}, but whether these are physically the right axioms in the first
place (or whether something more general is needed as it seems for $U_q(p)$) is
not clear: if $*$ does not respect the coproduct, it only means that the tensor
product of unitary representations is not necessarily unitary. This is caused
by the effective `interaction' due to the braiding $\Psi$ and is not
necessarily unphysical. See \cite{Ma:mec} for some initial results about
$*$-structures on braided groups and semidirect products.

Finally, let us mention that the above constructions can be mixed to give
various hybrid Poincar\'e-like ordinary Hopf algebras. For example, we can
bosonize both $V(R')$ and $V^*(R')$ (and not their duals as in Corollary~9) but
by the extended quantum enveloping algebra $\tilde H$ (and not by its
corresponding quantum function algebra $\tilde A$ as in Theorem~6 or
Proposition~7) to obtain ordinary Hopf algebras $V(R')\cocross \tilde H$ and
$V^*(R')\cocross \tilde H$. To describe these, it is convenient to write the
quantum enveloping algebra  $H$ (such as $U_q(so(1,3))$) with FRT-type
generators $\vecl^\pm=\{l^\pm{}^i{}_j\}$ as a quotient of a matrix bialgebra.
Here the FRT approach\cite{FRT:lie} is not limited to the standard
$R$-matrices, as we have demonstrated in \cite[Sec. 3]{Ma:qua}. Its extension
$\tilde H$ is simply the tensor product by the Hopf algebra $(\C\Z)^*=C(\Z)$
(functions on $\Z$) with formal quasitriangular structure\cite{Ma:csta}
$\CR(a,b)=\lambda^{-ab}$. There are several ways to describe this Hopf algebra,
for example as the group algebra of a circle via the Fourier transform. For our
purposes a convenient description is to choose a basis $\{\delta_a\}$ of
$\delta$-functions.  Also useful is powers of $\gamma=\sum_a\lambda^a\delta_a$.
Then the quasitriangular structure and the duality pairing with $\C\Z$ are
\eqn{RH}{\CR=\sum_{a,b}\lambda^{-{ab}}\delta_a\tens\delta_b=\sum_a\delta_a\tens
\gamma^{-a},\quad <\gamma^a,g^b>=\lambda^{ab}.}
Here $\Delta\delta_a=\sum_b\delta_b\tens\delta_{a-b}$ while
$\Delta\gamma=\gamma\tens\gamma$ is group-like.

\begin{propos} Denoting the generators of $V(R')$ by $p^i$, its bosonization by
the corresponding extended quantum enveloping algebra $\tilde H$ is as follows.
$V(R')\cocross \tilde H$ has cross relations and coproduct
\[ \gamma\vecp=\lambda^{-1}\vecp\gamma,\ \vecl_2^+
\vecp_1=\lambda^{-1}R_{12}^{-1}\vecp_1\vecl^+_2,\quad \vecl_2^-\vecp_1=\lambda
R_{21}\vecp_1\vecl^-_2,\quad \Delta\vecp=\vecp\tens 1+\vecl^-\gamma\tens
\vecp,\quad \Delta \vecl^\pm=\vecl^\pm\tens \vecl^\pm.\]
\end{propos}
\proof We use here the original form of the bosonization theorem\cite[Thm
4.1]{Ma:bos} by quantum enveloping algebras. The left action of $\vecl^\pm$ is
computed from the right coaction of $\tilde A$. For example, on $V(R')$ it is
$\vecl_2^+\la\vecp_1=<g^{-1}S\vect_1,\vecl^+_2>\vecp_1=<S\vect_1,\vecl^+_2>
\vecp_1=\lambda^{-1}R^{-1}_{12}\vecp_1$ and $\vecl_2^-\la \vecp_1=<gS\vect_1,
\vecl^-_2>\vecp_1=\lambda R_{21}\vecp_1$ which then immediately gives the
semidirect product algebra $(1\tens \vecl^\pm_2)(\vecp_1\tens 1)
=\sum \vecl^\pm_2\o\la\vecp_1\tens \vecl^\pm_2\t=(\vecl^\pm_2\la\vecp_1\tens 1)
(1\tens \vecl^\pm_2)$ in the form stated. The action of $\gamma$ is
$\gamma\la\vecp=<g^{-1}S\vect,\gamma>\vecp=<g^{-1},\gamma>\vecp=
\lambda^{-1}\vecp$ giving the semidirect product stated. For the coalgebra we
make the semidirect coproduct by the coaction $\beta(\vecp)=\CR_{21}\la\vecp
=\sum \CR\ut <\CR\uo,g^{-1}S\vect>\vecp=\vecl^-\sum\gamma^{-a}
<\delta_a,g^{-1}>
\tens\vecp=\vecl^-\gamma\tens\vecp$. Here we used $\sum\CR\ut<S\CR\uo,\vect>
=\vecl^-$ for the part in $H$ and (\ref{RH}) for the part in $C(\Z)$. Writing
this coaction as $\beta(\vecp)=\sum\vecp\bo\tens\vecp\bt$, the resulting
semidirect coproduct is $\Delta (\vecp\tens 1)=(\vecp\Bo\tens 1)(1\tens
\vecp\Bt\bo)\tens (\vecp\Bt\bt\tens 1)=\vecp\tens 1+\vecl^-\gamma\tens\vecp$
as stated. \endproof

There are analogous formulae for $V^*(R')\cocross \tilde H$. For example, we
have
\eqn{otherenvpoin}{\und{\R^{1,3}}\cocross \widetilde{U_q(so(1,3))}}
as a Hopf algebra that can be perhaps also be considered as some kind of hybrid
$q$-Poincare group. There are plenty of other possibilities. For example the
braided-tensor product of two braided-Hopf algebras is also a braided-Hopf
algebra, which can then likewise be bosonized to obtained an ordinary Hopf
algebra, for example $(\und{\R^{1,3}}\und\tens \und{\R^{1,3}})\cocross
\widetilde{U_q(so(1,3))}$ etc. The point we wish to make is that there is a
genuine and rich braided linear algebra\cite{Ma:lin} which can be used as
freely as familiar classical linear algebra (but remembering the braid
statistics) for constructions involving vectors and matrices.

\section{$q$-Lorentz Group as a Double Cross Product}

In the above we have assumed that the $q$-Lorentz group $SO_q(1,3)$ is given as
an $R$-matrix quantum group such as in \cite{CWSSW:lor}. We conclude by
mentioning that a version of this also has an abstract structure by which it is
built up from smaller factors. Here again the necessary abstract mathematical
constructions have been introduced previously by the
author\cite{Ma:phy}\cite{Ma:seq}, as a {\em double cross product}. This is a
generalization of a Hopf algebra semidirect product to the situation when both
Hopf algebras act on each other. Thus, if $H_1$ acts on $H_2$ from the left by
$\la$ and $H_2$ acts back from the right by $\ra$ in a compatible way, the
result is a double cross product Hopf algebra $H_1\bowtie H_2$. This coincides
with $H_1\tens H_2$ as a coalgebra but has product
\eqn{dcrossprod}{(a\tens b)(c\tens d)=\sum a(b\o\la c\o)\tens (b\t\ra
c\t)d,\quad a,c\in H_1,\ b,d\in H_2.}
where $\Delta a=\sum a\o\tens a\t$ etc. See \cite[Sec.
3.2]{Ma:phy}\cite{Ma:seq} for details. Here $H_1\bowtie H_2$ contains $H_i$ as
sub-Hopf algebras, with cross relations $(1\tens b)(c\tens 1)=\sum (b\o\la
c\o\tens 1)(1\tens b\t\ra c\t)$. We have introduced this construction some
years ago and shown that every Hopf algebra which factorizes into two sub-Hopf
algebras in a certain sense, is such a double cross product (this is a quantum
analog of a Manin triple).

For example,  we have shown that if $H_i$ act on the same space and as such
generate a bigger Hopf algebra, it is a double cross product. This was the
strategy behind the $q$-Lorentz group in \cite{CWSSW:lor} where the quantum
function algebra $SO_q(1,3)$ is realized as two copies of $SL_q(2)$ with matrix
generators $\vect,\vecs$ say and cross relations
\eqn{matcross}{R_{12}\vecs_1\vect_2=\vect_2\vecs_1R_{12}}
Here $R$ denotes the $SL_q(2)$ $R$-matrix. In fact, we had already studied
exactly this Hopf algebra or bialgebra in \cite[Thm 3.2]{Ma:seq} (previously to
\cite{CWSSW:lor}) where we obtained it for a general $R$-matrix as the double
cross product $A(R)\bowtie A(R)$ (or a Hopf algebra quotient of it) coming from
actions $\vecs_1\la\vect_2=R_{12}^{-1}\vect_2 R_{12}$ and
$\vecs_1\ra\vect_2=R^{-1}_{12}\vecs_1R_{12}$. We did not study its
$*$-structure or dual quasitriangular structure there.

This $A(R)\bowtie A(R)$ (including the $q$-Lorentz group in the form
$SL_q(2)\bowtie SL_q(2)$) is an example of the following abstract construction
for a dual-quasitriangular Hopf algebra $(A,\CR)$.

\begin{theorem} Let $(A,\CR)$ be a dual quasitriangular Hopf algebra. Then
there is a double cross product Hopf algebra $A\bowtie A$ built on $A\tens A$
as a coalgebra and with product
\[ (a\tens b)(c\tens d)=\sum \CR^{-1}(b\o\tens c\o)a c\t\tens b\t
d\CR(b\th\tens c\th),\qquad a,b,c,d\in A\]
This contains $A\tens 1, 1\tens A$ as sub-Hopf algebras. The actions giving
rise to this double cross product are induced by the left and right adjoint
coactions, turned into left and right actions by $\CR$ as
\[ b\la c=\sum \CR(Sb\tens c\o S c\th)c\t,\quad c\ra b=\sum
b\t\CR((Sb\o)b\th\tens c).\]
\end{theorem}
\proof This is a case of \cite[Thm 1.7]{Ma:seq} with $\CR\in (A\tens A)^*$ used
as an anti-self-duality pairing of $A$ with itself. We include here a direct
proof for completeness. Thus, $A$ coacts on $A$ by the right adjoint coaction
$b\mapsto \sum b\t\tens (Sb\o)b\th$, this becomes converted by $\CR$ to a right
action $\la$ (just as in the proof of Theorem~6 above and
\cite[Lemma~4.4]{Ma:mec}). Similarly $A$ coacts on $A$ by the left adjoint
coaction which becomes converted by $\CR^{-1}=\CR\circ(S\tens\id)$ to the left
action $\la$. One can check using the axioms of a dual-quasitriangular
structure that these actions are compatible in the way needed in \cite[Sec.
3.2]{Ma:phy}\cite{Ma:seq}. Hence we have a double cross product Hopf algebra.
The product (\ref{dcrossprod}) can then be computed as
\align{\nquad&&\nquad\sum b\o\la c\o\tens b\t\ra c\t=\sum \CR^{-1}(b\o\tens
c\o\o S c\o\th)c\o\t\tens b\t\t\CR((Sb\t\o)b\t\th\tens c\t)\\
&&=\sum \CR^{-1}(b\o\o\tens c\o\o)\CR^{-1}(b\o\t\tens S c\o\th)c\o\t\tens
b\t\t\CR(Sb\t\o\tens c\t\o)\CR(b\t\th\tens c\t\t)\\
&&=\CR^{-1}(b\o\tens c\o)\CR^{-1}(b\t\tens S c\th)c\t\tens
b_{(4)}\CR(Sb\th\tens c_{(4)})\CR(b_{(5)}\tens c_{(5)})}
where we used that the dual-quasitriangular structure $\CR$ is a skew
Hopf-bicharacter. This means that it is anti-multiplicative in its second
input, with the result that $\CR^{-1}$ is multiplicative in the sense
$\CR^{-1}(a\tens bc)=\sum \CR^{-1}(a\o\tens b)\CR^{-1}(a\t\tens c)$. Using this
again we can cancel $\CR^{-1}(b\t\tens Sc\th)\CR(Sb\th\tens c_{(4)})$ and
obtain the result stated in the theorem. \endproof

The resulting structure in Theorem~11 also works for $(A,\CR)$ a dual
quasitriangular bialgebra. This was the basis behind the example $A(R)\bowtie
A(R)$ in \cite{Ma:seq}. On the other hand, the abstract form in Theorem~11 also
makes clear a connection with an independent construction in \cite{ResSem:mat}
for the quantum double of a factorizable quantum group $H$ (such as
$U_q(sl_2))$) which gives $D(H)$ as isomorphic to $H\tens H$ as an algebra and
a doubly-twisted coalgebra. They denoted the latter $H\tens_{\CR}H$ (the
twisted square) and it is easy to see that it is exactly the dual Hopf algebra
to our double cross product $A\bowtie A$ with  $A$ dual to $H$. Thus, the dual
of the quantum double of a factorizable quantum group, is isomorphic to the
double cross product $A\bowtie A$. We have already shown in \cite[Example
4.6]{Ma:phy} that the quantum double $D(H)$ for any $H$ is a double cross
product $H^{*\rm op}\bowtie H$ by mutual coadjoint actions. Thus both quantum
doubles and their duals (in nice cases) are double cross products. In fact, the
formulae for $D(H)$ and $A\bowtie A$ are strictly analogous, $\CR$ playing the
role of a self-pairing as mentioned above.

In particular, the $q$-Lorentz group in the form $SL_q(2)\bowtie SL_q(2)$ is
also the dual of the quantum double of $U_q(sl_2)$ as proposed in
\cite{PodWor:def}. In algebraic terms the  latter is built on the algebra
$U_q(sl_2)\tens SL_q(2)$ with a doubly-twisted coproduct. The isomorphism
$SL_q(2)\bowtie SL_q(2)\to D(U_q(sl_2))^*$ is
\[ \vect\tens 1\mapsto S\vecl^-\tens \vect,\quad 1\tens \vecs\mapsto
S\vecl^+\tens \vect.\]
The same formulae hold for any dual-factorizable quantum function algebra $A$
of matrix type.

The abstract picture here is that for any $(A,\CR)$ the Hopf algebra $A\bowtie
A$ in Theorem~11 comes equipped with Hopf algebra maps $\pi_1:A\bowtie A\to
H^{\rm cop}$ (the dual $H$ of $A$ but with the opposite coalgebra) and
$\pi_2:A\bowtie A\to A$,
\eqn{proj}{\pi_1(a\tens b)=[(a\tens\id)(\CR)][(\id\tens b)(\CR^{-1})],\qquad
\pi_2(a\tens b)=ab}
and a Hopf algebra map $(\pi_1\tens\pi_2)\circ\Delta: A\bowtie A\to D(H)^*$
where $D(H)^*$ is built on $H^{\rm cop}\tens A$ as an algebra. In the
dual-factorizable case this becomes an isomorphism according to results in
\cite{ResSem:mat} and the projections become dual to the the usual inclusions
of $H$, $H^{*\rm op}$ in $D(H)$.
\cite{ResSem:mat} also proved that that twisted square is quasitriangular
(corresponding to the usual quasitriangular structure on $D(H)$). The dual
version of this is

\begin{propos} $A\bowtie A$ in Theorem~11 is dual quasitriangular with
\[ \CR((a\tens b)\tens (c\tens d))=\sum \CR^{-1}(d\tens (ab)\o)\CR((ab)\t\tens
c)\]
\end{propos}
\proof This is by direct computation using the axioms of a dual quasitriangular
structure and the structure of $A\bowtie A$ in Theorem~11. Note that the
canonical inclusions $A\subset A\bowtie A\supset A$ allow one to pull-back the
dual quasitriangular structure on $A\bowtie A$ to ones on each factor. The left
factor recovers its initial dual quasitriangular structure, while the right
factor recovers its inverse transpose one. \endproof

For example, we conclude that $A(R)\bowtie A(R)$ is dual-quasitriangular. The
value of the dual quasitriangular structure on the generators is
\eqn{Rbowtie}{\CR((\vect_1\tens\vecs_2)\tens(\vect_3\tens\vecs_4))
=R^{-1}_{41}R^{-1}_{42}R_{13}R_{23}}
where we used the multiplicativity properties of $\CR^{-1},\CR$ to write the
dual quasitriangular structure on $A\bowtie A$ equally well as a product of
four copies of $\CR$, and then evaluated. This also makes the connection with
\cite{ResSem:mat} transparent. On the other hand, because $A(R)\bowtie A(R)$ is
dual quasitriangular, we have in particular that
\eqn{lor}{\Lambda^K{}_B\Lambda^I{}_A R^A{}_J{}^B{}_L=R^I{}_A{}^K{}_B
\Lambda^A{}_J\Lambda^B{}_L}
where $I=(i_0,i_1)$, $J=(j_0,j_1)$ etc are multi-indices and
\eqn{lor-mat}{\Lambda^{(i_0,i_1)}{}_{(j_0,j_1)}=t^{i_0}{}_{j_0}\tens
s^{i_1}{}_{j_1},\quad
R^{(i_0,i_1)}{}_{(k_0,k_1)}{}^{(j_0,j_1)}{}_{(l_0,l_1)}
=R^{-1}{}^{k_1}{}_a{}^{i_0}{}_b R^{-1}{}^a{}_{l_1}{}^{i_1}{}_c
R^b{}_{j_0}{}^{k_0}{}_d R^c{}_{j_1}{}^d{}_{l_0}.}
{}From this we conclude at once that there is a bialgebra homomorphism
$A(R_{\bowtie})\to A(R)\bowtie A(R)$ given by $t^I{}_J\mapsto
\Lambda^I{}_J=t^{i_0}{}_{j_0}\tens s^{i_1}{}_{j_1}$, where $R_{\bowtie}$
denotes the composite $R$-matrix in (\ref{lor-mat}). For example, it means that
$SL_q(2)\bowtie SL_q(2)$ contains a quotient of the FRT bialgebra for the
corresponding composite $R$-matrix. Note that there are several $R$-matrices
that can be built up from four copies of an initial one, and the one in
(\ref{lor-mat}) appears different from the one used in \cite{CWSSW:lor} for a
similar purpose and obtained by twistor considerations. Another easy abstract
property of $A\bowtie A$, motivated by \cite{CWSSW:lor} is

\begin{propos}
If $A$ is a Hopf $*$-algebra over $\C$ and $\CR$ is if real type in the sense
$\overline{\CR(a\tens b)}=\CR(b^*\tens a^*)$ then $A\bowtie A$ is a Hopf
$*$-algebra with $(a\tens b)^*=b^*\tens a^*$.
\end{propos}
\proof This is by direct computation. Thus,
\align{((a\tens b)(c\tens d))^*&&=\sum \overline{\CR(Sb\o\tens c\o)}(a c\t\tens
b\t d)^*\overline{\CR(b\th\tens c\th)}\\
&&=\CR(c^*\o\tens S^{-1}(b^*\o)) d^* b^*\t \tens c^*\t a^* \CR(c^*\th\tens
b^*\th)=(d^*\tens c^*)(b^*\tens a^*)}
so that $*$ on $A\bowtie A$ is an anti-algebra homomorphism. We used here the
assumed reality property of $\CR$ and the various definitions. The last
equality uses the invariance $\CR\circ(S\tens S)=\CR$ to write $\CR(c^*\o\tens
S^{-1}(b^*\o))=\CR(Sc^*\o\tens b^*\o)$. That $*$ commutes with the tensor
coalgebra structure of $A\bowtie A$ and the other axioms for a Hopf $*$-algebra
are straightforward. \endproof

For example, $SU_q(2)$ at real $q$ obeys the reality condition for its
$\CR$\cite{Ma:mec} and the last proposition gives  the $*$-structure on
$SU_q(2)\bowtie SU_q(2)$ as
\eqn{lor-*}{t^i{}_j{}^*=Ss^j{}_i,\qquad  s^i{}_j{}^*=St^j{}_i,}
which is of the form used in \cite{CWSSW:lor}.

In this way, we see that several constructions in the literature regarding the
$q$-Lorentz group are related through the theory of double cross products.
Other complicated quantum groups can surely likewise be understood by means of
such semidirect or double-semidirect product constructions. For example, the
Poincar\'e groups in \cite{LRNT:def} can perhaps be understood in terms of the
general extension theory of Hopf algebras and bicrossproducts\cite{Ma:seq}.
Also, there are signs that the twistor construction motivating \cite{CWSSW:lor}
can be understood in terms of braided tensor products as in \cite{Ma:lin}.
These are topics for further work.

\new{Acknowledgements} I would like to thank M. Schlieker and coworkers present
at the 2nd Max Born Symposium, Wroclaw, Poland 1992, for explaining to me their
work and drawing my attention to \cite{SWW:inh}. The main results above were
obtained at the symposium in response to this.

\end{document}